\documentclass[aps,prl,twocolumn,superscriptaddress]{revtex4}
\usepackage{amssymb}
\usepackage{graphicx}
\usepackage{amsmath}

\begin{document}

\title{Charge Pumping of Interacting Fermion Atoms in the Synthetic Dimension}

\author{Tian-Sheng Zeng}
\affiliation{School of Physics, Peking University, Beijing 100871, China}
\author{Ce Wang}
\affiliation{Institute for Advanced Study, Tsinghua University, Beijing, 100084, China}
\author{Hui Zhai}
\affiliation{Institute for Advanced Study, Tsinghua University, Beijing, 100084, China}

\date{\today}

\begin{abstract}

Recently it has been proposed and experimentally demonstrated that a spin-orbit coupled multi-component gas in 1d lattice can be viewed as spinless gas in a synthetic $2$d lattice with a magnetic flux. In this letter we consider interaction effect of such a Fermi gas, and propose signatures in charge pumping experiment, which can be easily realized in this setting. Using $1/3$ filling of the lowest 2d band as an example, in strongly interacting regime, we show that the charge pumping value gradually approaches a universal fractional value for large spin component and low filling of 1d lattice, indicating a fractional quantum Hall type behavior; while the charge pumping value is zero if the 1d lattice filling is commensurate, indicating a Mott insulator behavior. The charge-density-wave order is also discussed. 

\end{abstract}

\maketitle

High spin quantum gas is a unique system of cold atom physics. The spin of atoms ranges from hyperfine $F=1$ of alkali atoms like Rb and Na, $F=9/2$ of K atoms to nuclear spin with $SU(W)$ symmetry of alkali-earth-(like) atoms such as Yb and Sr, where $W$ can be as large as $\sim 10$. Recently it emerges an interesting idea to use the internal spin degrees of freedom as another dimension, named as ``synthetic dimension", which naturally extends a $D$-dimensional system into a $(D+1)$-dimension one \cite{extraD,syntheticD}. In a 1d lattice system, by applying two counter propagating Raman beams to couple different spin states, one can create a magnetic flux lattice in synthetic 2d \cite{syntheticD}. This proposal requires a minimum amount of laser light and therefore minimizes heating from spontaneous emission. It also gives rise to a sharp edge in the synthetic dimension, which can help to visualize edge states. Very recently, two experimental groups have implemented this scheme, in Rb atom \cite{NIST} and in Yb atom \cite{LENS}, respectively, and chiral edge states have been observed, for non-interacting (or weakly interacting) bosons \cite{NIST} and fermions \cite{LENS}, respectively. Moreover, it is possible to create more exotic nontrivial geometry \cite{nontrivial}.   

\begin{figure}[t]
\includegraphics[width=3.4 in]
{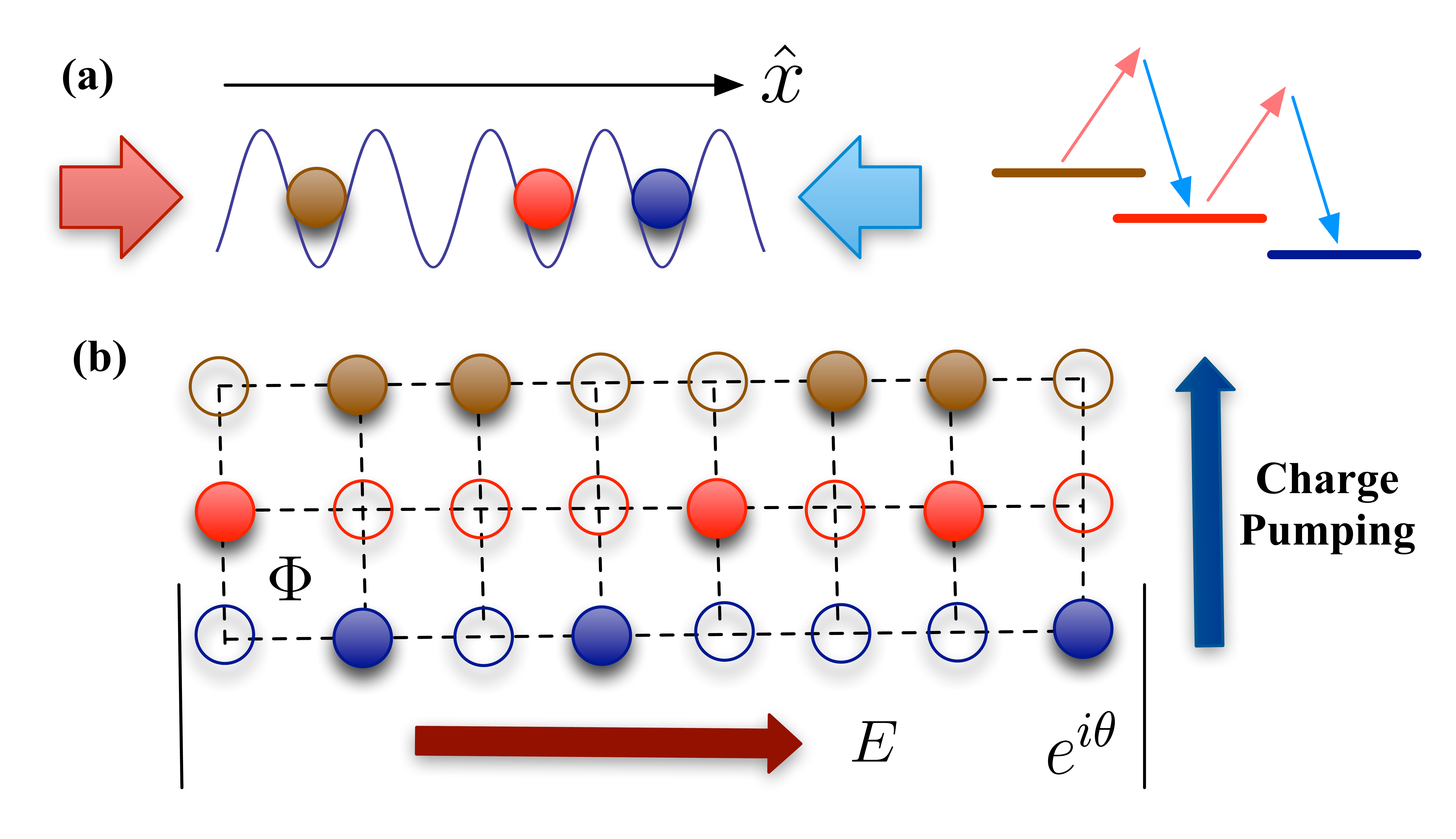}
\caption{Illustration of two different pictures of this system and the idea of charge pumping. (a) Two Raman beams along $\hat{x}$ are applied to a multi-component quantum gases in optical lattices. The Raman beams couple different spin states and generate a coupling between spin and momentum $k_x$. (b) Different spin components are viewed as another dimension, the system is mapped into a 2d spinless particle in a magnetic field. By applying an electric field along the physical dimension $\hat{x}$, it generates a charge pumping along the synthetic dimension, which can be detected by measuring spin populations. }
\label{model}
\end{figure}

The experimental setup and basic idea of synthetic dimension are briefly illustrated in Fig. \ref{model}. For instance, two Raman beams with $\pi$ and $\sigma$ polarization can couple spin state $|m\rangle$ to $|m\pm 1\rangle$, where $m$ can run from $-F$ to $F$ with totally $W=2F+1$ components. The Raman coupling has a spatial dependent phase factor $e^{i2k_\text{R}x}$, where $k_\text{R}$ is the recoil momentum of Raman laser. We consider the situation $\gamma=2k_\text{R}a=2\pi/q$ ($a$ is lattice spacing). The single particle Hamiltonian is therefore written as
\begin{equation}
\hat{H}_0=\sum\limits_{j,m}(-t \hat{c}^\dag_{j+1,m}\hat{c}_{j,m}+\Omega e^{-i\gamma j}\hat{c}^\dag_{j,m-1}\hat{c}_{j,m}+\text{h.c.}) \label{H0}
\end{equation}
where $j$ labels the site along the physical dimension $\hat{x}$ and $m$ labels internal spin components. $t$ is hopping along $\hat{x}$ and $\Omega$ is Raman coupling strength. There are two different but equivalent pictures of this single-particle Hamiltonian

(a) A 1d system of high spin atoms with spin-orbit coupling: by applying a spin and site dependent rotation $\hat{c}_{j,m}\rightarrow e^{i\gamma j m}\hat{c}_{j,m}$, the Hamiltonian Eq. \ref{H0} becomes
\begin{equation}
\hat{H}_0=\sum\limits_{j,m}(-t e^{-i\gamma m}\hat{c}^\dag_{j+1,m}\hat{c}_{j,m}+\Omega \hat{c}^\dag_{j,m-1}\hat{c}_{j,m}+\text{h.c.}) \label{HSO}
\end{equation}
The spin dependent hopping term, together with a constant spin flipping coupling, gives rise to the spin-orbit coupling effect, which have been extensively discussed for spin-$1/2$ case in continuum in the past a few years \cite{exp,Jing, MIT,review}.

(b) A 2d system with magnetic flux: If we view $m$ as another synthetic dimension, say, labelled by $\hat{y}$, Eq. \ref{H0} represents a situation that spinless atoms hops in a 2d space, with finite number ($W$) of chains and open boundary condition along $\hat{y}$. More importantly, hopping along a close loop around a plaquette accumulates a phase of $e^{i\gamma}$, that is equivalent to say, each plaquette has a flux of $\Phi_0/q$ ($\Phi_0$ is magnetic flux unit), which is gauge coupled to motion of atoms. 

These two equivalent pictures build up an intriguing connection between spin-orbit coupled high spin particles in 1d and spinless but charged particles in a 2d ladder geometry with magnetic flux. For instance, at single-particle level, the chiral edge current from the picture (b), as observed in Ref. \cite{NIST,LENS}, is equivalent to spin-momentum locking effect from picture (a), as observed in spin-$1/2$ case \cite{exp,Jing,MIT}.  In this letter we aim at studying effect of repulsive interaction in this system \cite{attractive}. For 2d electron gas in a magnetic field, it is known that fractional quantum Hall (FQH) state will emerge with strong repulsive interactions. However, in this system there are a few important differences that worth emphasizing first:

1. In the synthetic dimension, the system always only has finite number of chains. Though it is possible to create a periodic boundary condition in synthetic dimension with some complicated laser setting, in the most simple and natural setup, the system has an open boundary condition in synthetic dimension. Thus, the finite size effect can be significant. 

2. Along the physical dimension, the interaction is on-site and short-ranged. While in the synthetic dimension, the interaction is long-ranged. Let us consider a $SU(W)$ invariant interaction \cite{footnote},
\begin{equation}
\hat{H}_\text{int}=U\sum\limits_{j,m\neq m^\prime}\hat{n}_{j,m}\hat{n}_{j,m^\prime}, \label{Hint}
\end{equation} 
atoms in any two sites along the synthetic dimension, despite of their separation, interact with the same interaction strength. In another word, in the 2d lattice the interaction is very anisotropic. 

3. For normal FQH effect, the only relevant parameter $\nu$ is the ratio between fermion number to flux number. In our case, it will be
\begin{equation}
\nu=\frac{N}{N_\text{flux}}=\frac{N}{WL/q}=\frac{Nq}{WL},
\end{equation}
where $N$ is total number of fermions, $L$ is the number of sites along $\hat{x}$-direction. However, from the picture (a) that our system is one-dimensional, it is natural to introduce another filling factor 
\begin{equation}
\nu_{\text{1d}}=\frac{N}{L},
\end{equation}
and if $\nu_{\text{1d}}$ is an integer, one may expect a trivial Mott insulator rather than a FQH state when interaction $U$ is sufficiently large. Thus, when $\nu$ is fixed, we still have two other tunable parameters, i.e. $\nu_{\text{1d}}$ and $W$. 

Hereafter we shall fix $\nu=1/3$ as a typical example. Given those differences mentioned above, one may wonder whether one will still have a FQH-type behavior under strong repulsive interactions. Beside, whether there is an effective scheme to detect such a state in this cold atom setting. The rest of this paper is devoted to answer this question. 

\textit{Charge Pumping.} To reveal the interaction effect, we propose to perform a charge pumping experiment utilizing the advantage of synthetic dimension. Let us consider applying an electric field along $\hat{x}$ direction. In our numerical calculation below, this is implemented by a periodic boundary condition in $\hat{x}$ direction, i.e. $\Psi(L)=e^{i\theta}\Psi(0)$. In cold atom experiment, this can be realized by, for instance, applying a field gradient or moving lattice with a constant velocity.  Charge pumping here means charge transfer along the synthetic dimension \cite{Thouless}, as schematized in Fig. \ref{model}.  Defining the charge center as  
\begin{equation}
Y=\frac{1}{W}\sum_{j,m}\langle \hat{n}_{j,m}\rangle m, \label{Y}
\end{equation}
the charge pumping after inserting one flux is given by
\begin{equation}
Q=Y(\theta=2\pi)-Y(\theta=0). \label{Q}
\end{equation} 
If it is in real space, detecting charge transfer requires in-situ image, while charge transfer in synthetic dimension means changing of spin population, which can be much easily detected from the Stern-Gerlach experiment in the time-of-flight image \cite{Cooper,Shuai}.  

\begin{figure}[tbp]
\includegraphics[width=3.3 in]
{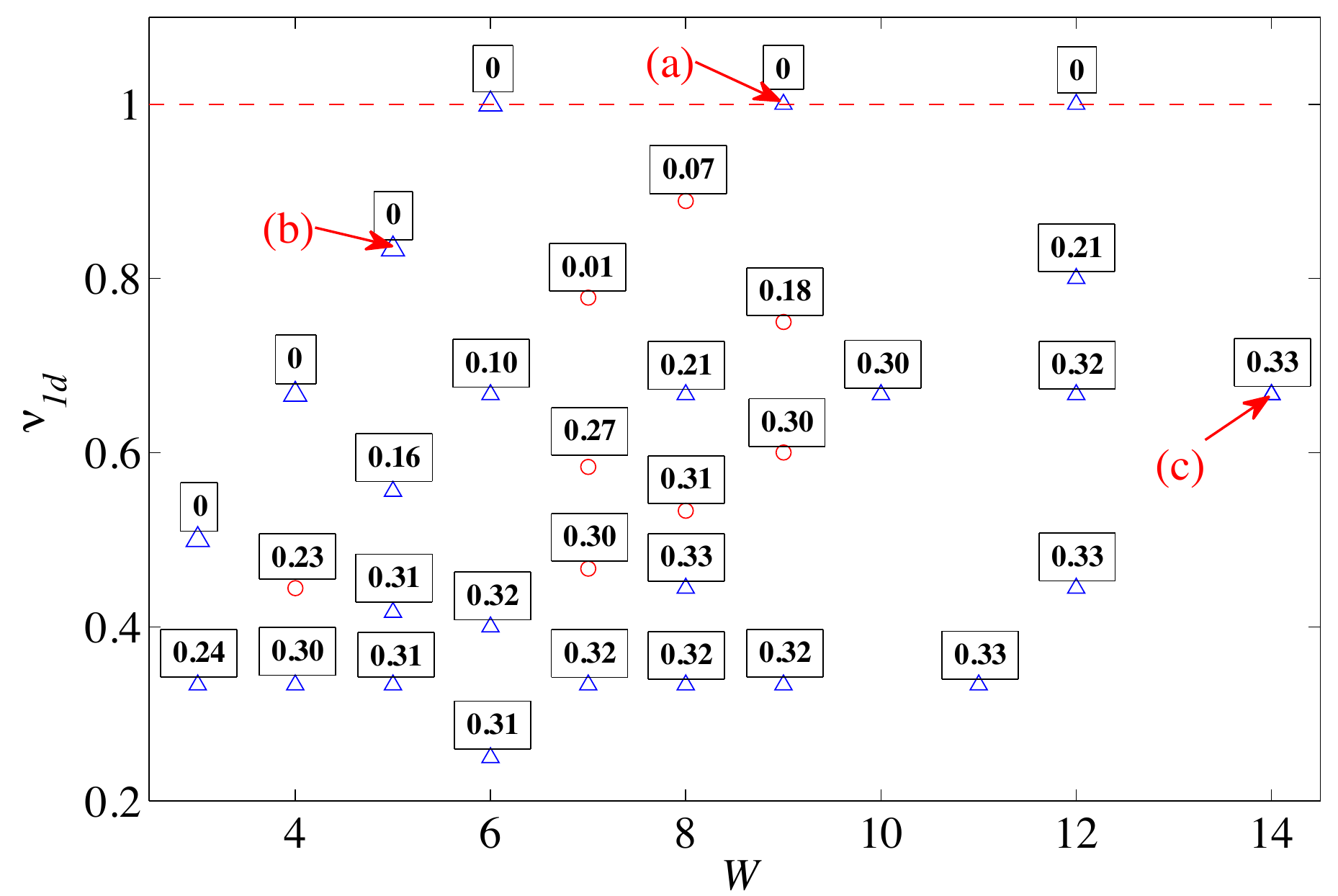}
\caption{Charge pumping for different $\nu_\text{1d}$ and $W$. Numbers in box are charge pumping $Q$ after insertion of one flux. Blue triangles are points calculated by ED and red circles are points calculated by DMRG, with $\nu$ fixed at $1/3$, $\Omega=t$ and $U=6t$. (a-c) marked three cases where spectral flow will be shown in Fig. \ref{pumping}.}
\label{phasediagram}
\end{figure}

In Fig. \ref{phasediagram}, we present the charge pumping value $Q$ for various $\nu_{\text{1d}}$ and $W$, with $\nu$ fixed at $1/3$. This result is obtained by numerically solving the many-body wave-functions with Hamiltonian $\hat{H}=\hat{H}_0+\hat{H}_\text{int}$, either by exact diagnolization (ED) or density matrix renormalization group (DMRG) methods. For ED the maximum number of particle is six and the dimension of Hilbert space is of the order of $3\times 10^7$. For DMRG, the maximum number of particle is ten, and the truncation error is of the order of $10^{-7}$. Each eigenstate has a well-defined quantum number $K$ that is the center-of-mass momentum along $\hat{x}$. We also plot how these eigenstates evolve under the changing of $\theta$, as shown in Fig. 3, and for ground state, we calculate $Y$ as a function of $\theta$ with Eq. \ref{Y} and deduce $Q$ with Eq. \ref{Q}. 

\begin{figure}[tbp]
\includegraphics[width=3.4 in]
{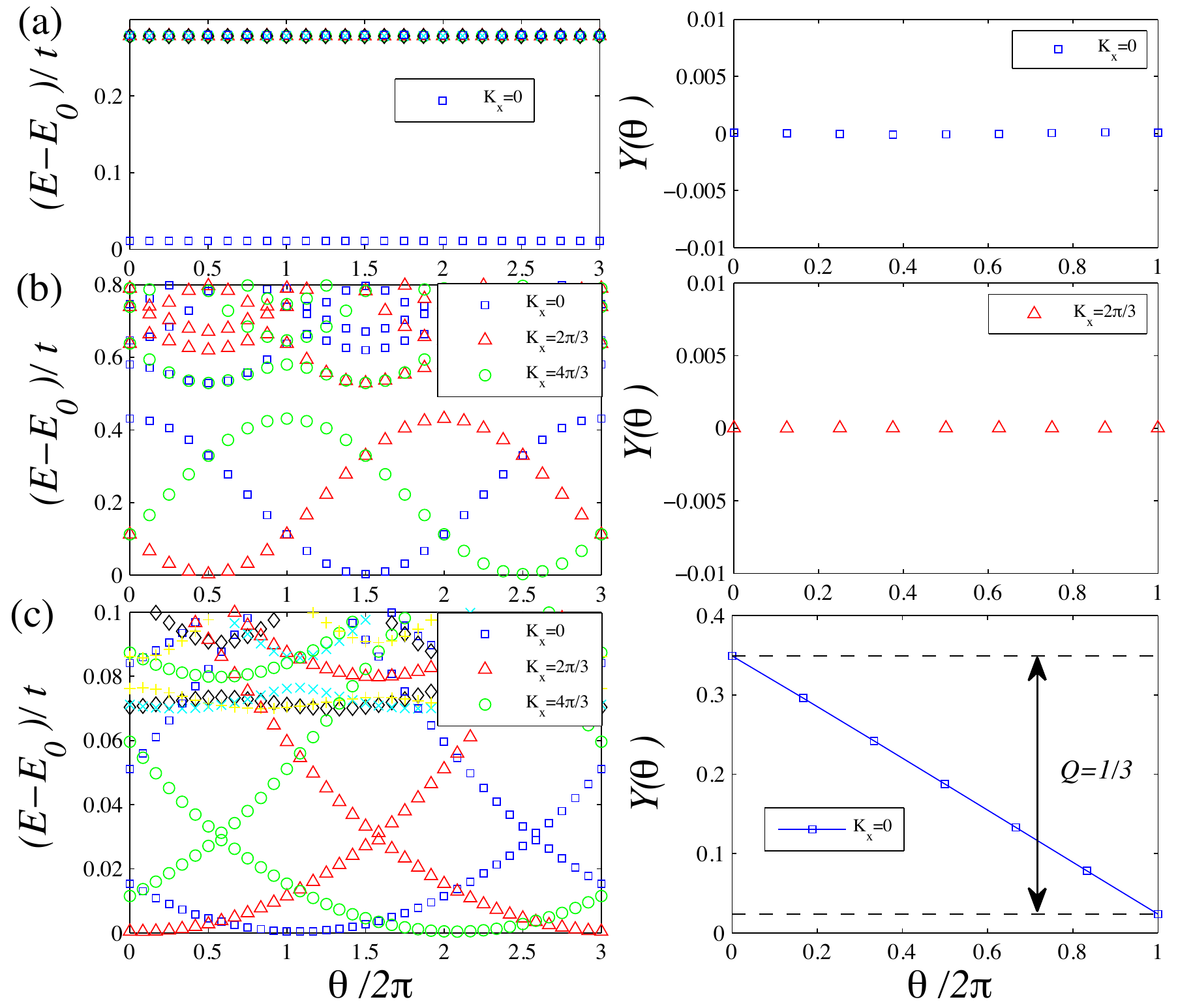}
\caption{Left column: Spectral flow under the insertion of flux (i.e. changing periodic boundary condition $\theta$ from zero to $6\pi$.) Right column: charge center $Y$ as a function of $\theta$. (a-c) correspond to different $W$ and $\nu_\text{1d}$ as marked in Fig. \ref{phasediagram}. (a) $W=9$, $\nu_\text{1d}=1$ ($N=L=5$); (b) $W=5$, $\nu_\text{1d}=5/6$ ($N=5$, $L=6$) and (c) $W=14$, $\nu_\text{1d}=2/3$ ($N=4$, $L=6$). All these cases are calculated by ED.}
\label{pumping}
\end{figure}

We find following features: (i) For $\nu_{\text{1d}}=1$, $Q$ is identically zero for all $W$. In this case there is always a unique ground state which will not interchange with other states under flux insertion, as shown in Fig. 3(a). This Mott insulator phase with commensurate $\nu_\text{1d}$ is due to one-dimensional nature of the system, or equivalently to say, anisotropic nature of interaction in picture (b). (ii) Since here we consider $q$ is an integer, the smallest value for $q$ is $q=2$. With $q=2$ and $\nu=1/3$, $\nu_\text{1d}=\nu W/q$ can at most be $W/6$. We note that for $q=2$, the Hofstadter spectrum exhibits Dirac cone instead of fully gapped band, and the lowest band does not have well defined Chern-number. Therefore, for $\nu_\text{1d}=W/6$ (with $W\leqslant 6$) we also find $Q=0$.  A typical spectral flow and charge pumping is shown in Fig. 3(b). (iii) For small W, or for larger $W$ but $\nu_\text{1d}$ closer to unity, though $Q$ is generally non-zero, it takes non-universal values. This fluctuating $Q$ indicates some Fermi-liquid type states \cite{Sheng}. (iv) For large $W$ and smaller $\nu_\text{1d}$, $Q$ gradually approaches a universal fractional value of $1/3$. A typical spectral flow under flux insertion is shown in Fig. 3(c). One can see that there are three low-lying states (though not exactly degenerate) exchange one with the other as $\theta$ increases, and the spectrum recovers itself only after $\theta$ changes $6\pi$.  These features are consistent with a FQH effect. This is because for large $W$, the finite size effect in synthetic dimension becomes insignificant, and for smaller $\nu_\text{1d}$ away from commensurate filling, the lattice effect also becomes weaker. Previously, it has been shown by Luttinger liquid theory that for continuum models, fractional state can emerge in a one-dimensional system with spin-orbit coupling \cite{LL}.  

\begin{figure}[tbp]
\includegraphics[width=3.4 in]
{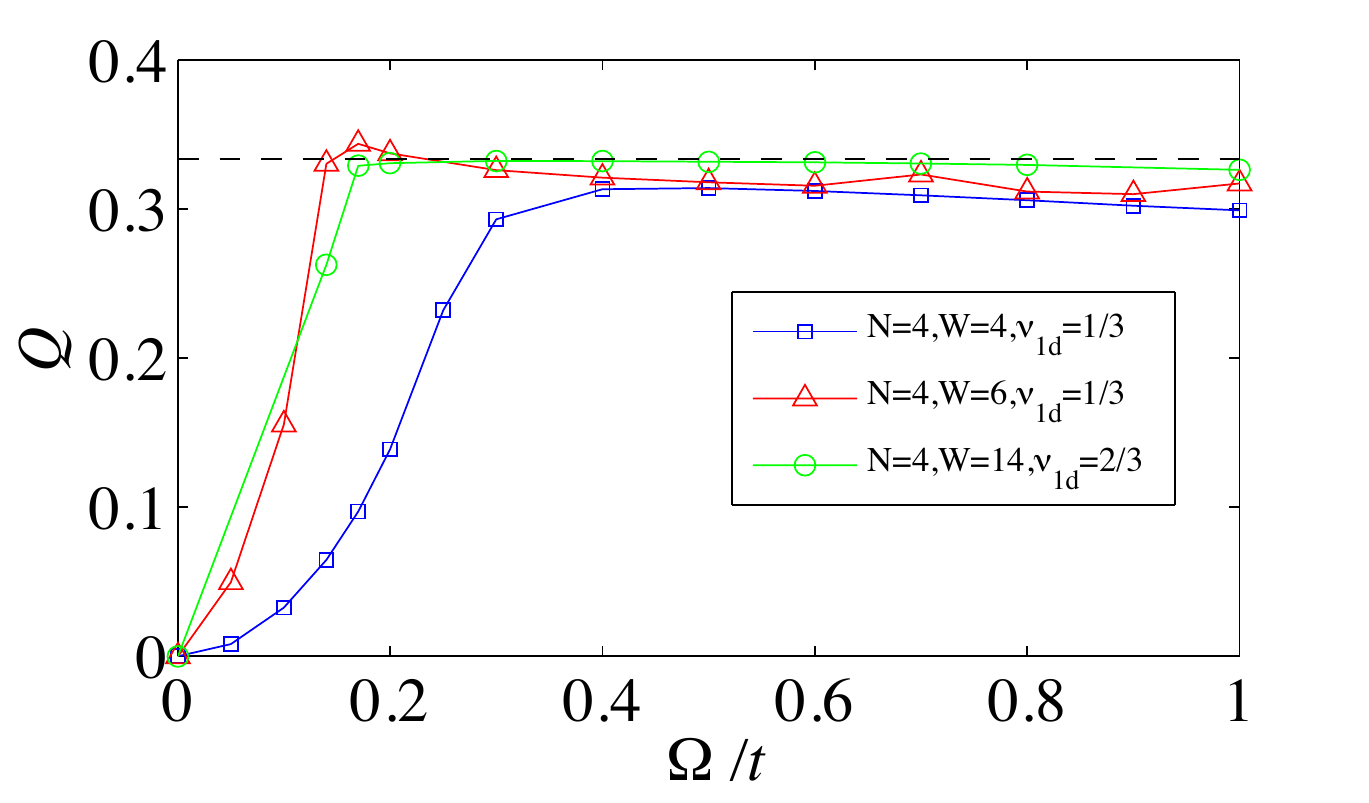}
\caption{Charge pumping $Q$ as a function of Raman coupling strength $\Omega/t$. The dashed line is $1/3$. }
\label{pumpingOmega}
\end{figure}

We also study the charge pumping value $Q$ as a function of Raman coupling strength $\Omega$, as shown in Fig. \ref{pumpingOmega}. As the synthetic magnetic field is resulted from Raman coupling, one naturally expects that $Q$ will vanish as $\Omega\rightarrow 0$. Indeed, we show in Fig. \ref{pumpingOmega} that when $\Omega/t$ is smaller than certain value, $Q$ starts to derivates from $1/3$ and fast drops to zero. This feature is particularly clear for large $W$ (e.g. green points for $W=14$ in Fig. \ref{pumpingOmega}). We have also looked at $Q$ for smaller $U/t$ and find when $U<t$,  $Q$ also takes fluctuating non-universal values.  

\begin{figure}[bp]
\includegraphics[width=3.4 in]
{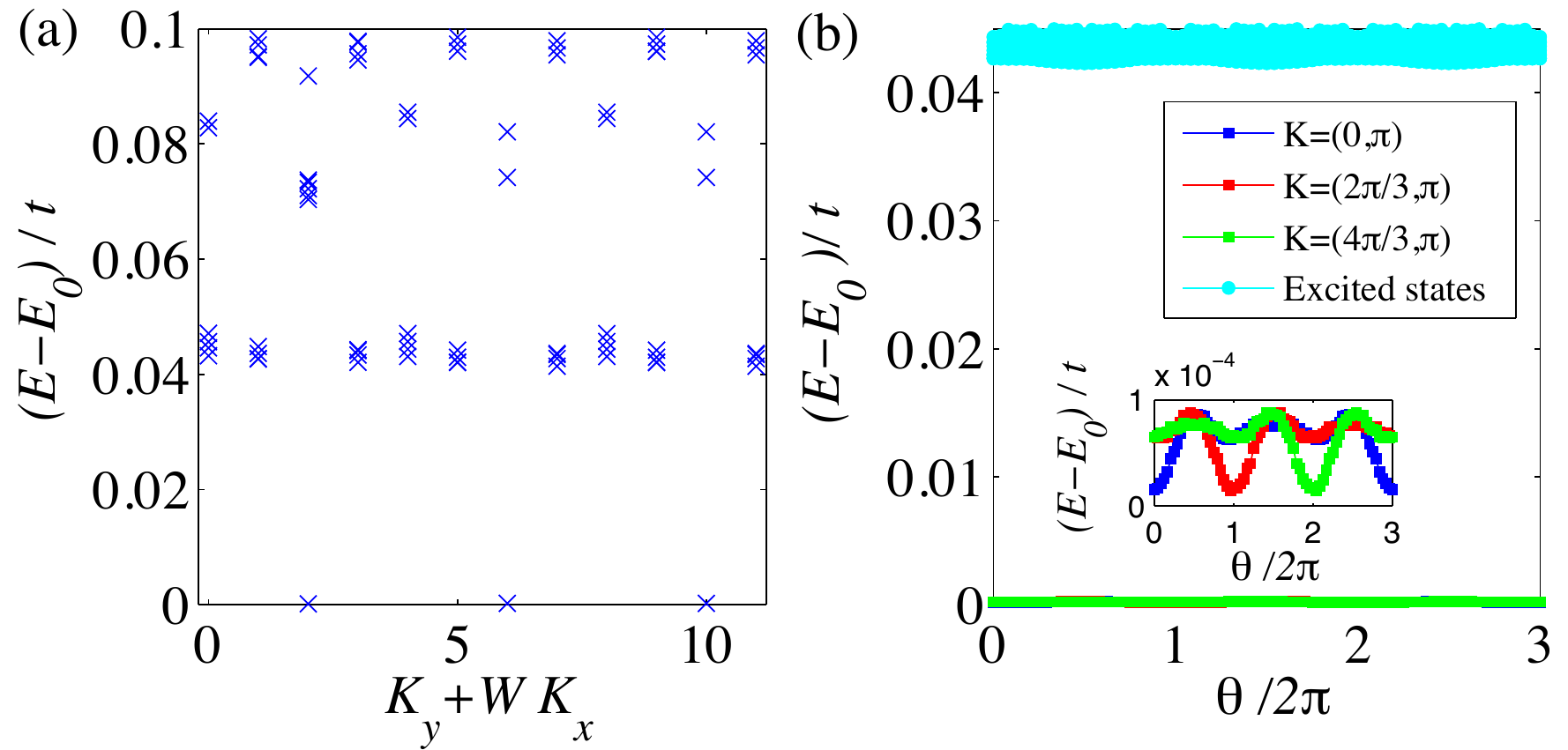}
\caption{Energy diagram with periodic boundary condition in synthetic dimension. (a) Energy levels for different momentum $K_y+WK_x$. (b) Energy of the lowest three states as a function of $\theta$. Here $W=4$, $\nu_\text{1d}=1/3$ ($N=4$, $L=12$). This case is also calculated by ED. }
\label{periodic}
\end{figure}

\textit{Periodic Boundary Condition in Synthetic Dimension.} We also find that if one applies a more involved laser setting to achieve a periodic boundary condition along the synthetic dimension, it will greatly help to stabilize a fractional state. For instance, for the cases with $W=4$ we presented in Fig. \ref{phasediagram}, we do not find accurate fractional charge pumping. While when we apply a periodic boundary condition in synthetic dimension, we show the energy level for different momentum (labelled by $K_y+WK_x$ as now both $K_x$ and $K_y$ are good quantum numbers) in Fig. \ref{periodic}. We find a very accurate three-fold degeneracy, with energy splitting smaller than $10^{-4}t$, and these states are separated from other excited states by a gap $\sim 0.04t$. The total momentum of these three states are also consistent with generalized Pauli-exclusion principle analysis \cite{Bernevig}. Moreover, these states exchange one and the other under the flux insertion, and do not intersect with other excited states, as shown in Fig. \ref{periodic}(b). By calculating Berry curvature with twisted boundary conditions in both physical and synthetic dimensions \cite{Sheng}, we numerically  find that their many-body Chern number $\mathcal{C}_1=\mathcal{C}_2=\mathcal{C}_3=0.333$. 

\begin{figure}[tbp]
\includegraphics[width=3.3 in]
{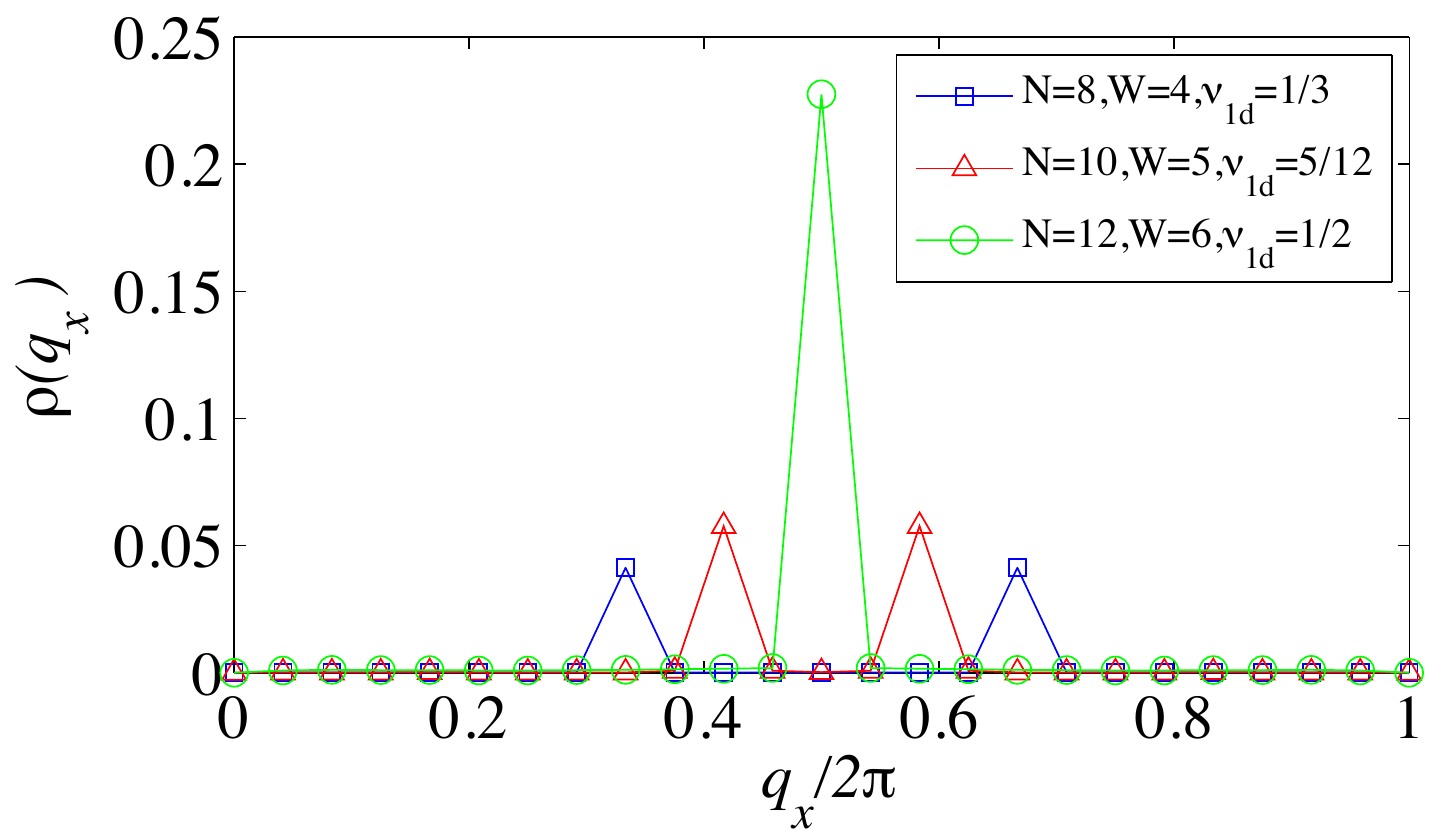}
\caption{Fourier transformation of density $\rho(q_x)$ for different $W$ and $\nu_\text{1d}$ }
\label{density}
\end{figure}

\textit{Density Order.} Finally we look at real-space charge density-order. We consider the onsite total density $\rho_j=\sum_m\langle \hat{n}_{j,m}\rangle$. In Fig. \ref{density}, we plot the Fourier transform of $\rho_i$ as $\rho(q_x)=(1/L)\sum_{j}(\rho_j-\bar{\rho}) e^{iq_x j}$ ($\bar{\rho}=\nu_\text{1d}$ is the average density). $\rho(q_x)$ shows a clear peak at $q_x/(2\pi)=\nu_\text{1d}$ and $q_x/(2\pi)=1-\nu_\text{1d}$. This feature exists for both open and periodic boundary condition in synthetic dimension. Similar situation has also been found in several other models \cite{CDW}. This is reminiscent of usual FQH state in the thin torus limit \cite{DHL}.

\textit{Conclusion.} In summary, we have studied interaction effects  in the synthetic dimension picture of high spin lattice Fermi gases with Raman-coupling induced spin-orbit coupling. Our studies are mainly focused on signatures of charge pumping experiment, which becomes much easier in this setting, as charging pumping along the synthetic dimension can be visualized by measuring spin population. For fixed $\nu=1/3$ case, we investigated how the charge pumping value depends on number of spin component $W$, fermion density $\nu_\text{1d}$, and Raman coupling $\Omega/t$. We conclude that a universal fractional charge pumping $Q=1/3$ is favorable for $W\gg 1$, $\nu_\text{1d}\ll 1$, $\Omega/t \sim 1$ and with strong interactions. Similar results have also been obtained for strongly interacting bosons.

\textit{Acknowledgment}:  We would like to thank Hong Yao for helpful discussion. This work is supported by Tsinghua University
Initiative Scientific Research Program, NSFC Grant No. 11174176, No. 11325418 and NKBRSFC
under Grant No. 2011CB921500.

\textit{Note Added.} When finishing up the manuscript, we become aware of another paper Ref. \cite{Mazza} in which the same system is studied by DMRG. Charge density wave order is also discussed in this paper.


\begin{thebibliography}{99}
\bibitem{footnote}
For alkali-earth atom, the interaction is $SU(W)$ invariant. For alkali atom, this is also a very good approximation since the spin-dependent interaction is usually quite small and the spin mixing collision takes place in a much longer time scale than the time scale for synthetic magnetic field and $SU(W)$ invariant part of density interaction.
\end{thebibliography}
\end{document}